\documentclass[aps,prl,twocolumn,floats,showpacs,superscriptaddress]{revtex4-1}
\usepackage{graphicx}
\usepackage{xcolor}
\title{Results}
\date{\today}

\usepackage{amsmath}
\usepackage{hyperref}
\usepackage{lineno}
\begin{document}
	\title{Co-evolution of Cooperation and Epidemic Spreading }	
	
	\author{Mehran Noori}
	\affiliation{Physics Department, Institute for Advanced Studies in Basic Sciences (IASBS), Zanjan 45137-66731, Iran.}
	
	\author{Nahid Azimi-Tafreshi}
	\email{nahid.azimi@iasbs.ac.ir}
	\affiliation{Physics Department, Institute for Advanced Studies in Basic Sciences (IASBS), Zanjan 45137-66731, Iran.}

	\author{Mohammad Salahshour}
	\email{msalahshour@ab.mpg.de}
	\affiliation{Max Planck Institute of Animal Behavior, Konstanz, Germany.}
	\affiliation{Centre for the Advanced Study of Collective Behaviour, University of Konstanz, Konstanz, Germany.}
	\affiliation{Department of Biology, University of Konstanz, Konstanz, Germany.}

	\begin{abstract}
		People's cooperation in adopting protective measures is effective in epidemic control and creates herd immunity as a public good. Similarly, the presence of an epidemic is a driving factor for the formation and improvement of cooperation. Here, we study the coevolution of epidemic dynamics and the public goods game as a paradigm of cooperation dynamics.
		Using simulations and a mean-field description, we show that the interaction of the evolutionary dynamics of cooperation and epidemic spreading leads to a feedback loop between the two dynamics. The higher disease transmission rate can promote cooperation and lead to the reduction of disease spread; in turn, the higher benefit of cooperation leads to the more efficient control of epidemics.
		Additionally, our work shows that a higher altruistic effect of cooperation in controlling the disease can only be detrimental to the evolution of cooperation and disease control. Rather, individuals' choices in adopting costly preventive measures are predominantly driven by the self-interested effect of such measures in reducing the probability of getting infected.
		\end{abstract}	
	\maketitle
	\section{Introduction}
	
	Epidemic spreading has been subject to extensive research \cite{bogua2003,wang2024,koher2019,zhou2019,moez2006} due to its crucial impact on public health and societal stability \cite{newman2002,pastor2015,wang2017}. A critical goal of scientific inquiry on epidemic dynamics is achieving a better understanding of the mechanisms by which epidemic outbreaks may be controlled \cite{wang2016,Han,Claudio,Bavel}. Behavioral responses of individuals to epidemic spreading, such as adopting preventive measures, for instance, vaccination \cite{pastor2002,schneider2011,khanjani2020,wang2016} and social distancing \cite{gosak2021community,gosak2021endogenous} can be effective in controlling the spread of disease. In this regard, the computational models played an important role in supporting the policy makers during the epidemic, in particular the COVID-19 Scenario Modeling Hub (SMH) provided modeling evidence to guide decisions in the evolving pandemic \cite{SMH1,SMH2}. 
	
	Most preventive measures are associated with social and financial costs for the individual. Yet, they are beneficial for the community as a whole by reducing the disease spread. As such, these preventive measures qualify as public goods \cite{morison2024,dees2018,yong2021,soltan2020}--- common goods which are costly for the individual but beneficial for the group--- and their adoption by the individuals can be studied using game theoretic methods \cite{tanimoto2015wp,chang2020ug,huang2022qh,amaral2021}. A focal point of interest in this regard is the feedback loop between individuals' decision-making and epidemic spreading, which can lead to a rich coevolutionary dynamics of individuals' behavior and epidemic spreading \cite{hamilton,wang2020uc,wang2015coupled,wang2020vaccination,khanjanianpak2022ap}.
	
	Despite the importance of behavioral responses and their classification as public goods, and despite the fact that public goods has served as a focal point of research on human cooperation in evolutionary game theory \cite{haurent2002,hauert2003wc,hauert2008ecological,santos2008social,xianbin2010,perc2017,Zhu2024}, little attention has been paid to the interaction between epidemic dynamics and the provision of these public goods. 
	In this framework, the COVID-19 situation has been studied as a public goods dilemma, where people who neglect safety precautions act as free riders, and the government needs to prepare the communities by encouraging individuals to cooperate \cite{yong2021, Tim2020}. Also, Ref.~\cite{soltan2020} studies the role of an incubation period in an epidemic outbreak by considering the SIRV epidemic model with four categories for disease status, and a public good setting with payoff functions based on the cost of living in a group and the individual cost related to the epidemic state of each player. However, the coevolutionary dynamics of epidemics and public goods—-- that is, the question of how the spread of disease may influence cooperative behavior and how, in turn, cooperation may impact the course of the epidemic, remains an underexplored area in the literature.
	
	In this study, we aim to bridge this gap by investigating the interplay between epidemic dynamics and public goods. We introduce a coevolutionary model of public good providing and epidemic spreading. Individuals are either susceptible or infected. In addition, they can be cooperator or defector. Cooperation, while costly, provides collective benefit both via the public goods game (PGG) and through reduced risk of infection. Using the mean-field approximations and numerical simulations in well-mixed and structured populations, we explore the interplay between epidemic spreading and the evolution of cooperation.
	
	The article is organized as follows. In the next section, we define the dynamics of the model. In \ref{methods} Methods section  we derive the time evolution of populations in the mean field approximation and also discuss the details of the numerical simulations. We present the results in a well-mixed population and also the case where the population is arranged on a square lattice in the \ref{results}Results section. The paper is concluded in the \ref{conclusion}Conclusion section.

\section{The Model}
	\label{definition}
	We consider a population of $N$ individuals playing a public goods game and subject to an epidemic spreading. The disease spreading is modeled using the susceptible-infected-susceptible (SIS) epidemic model. Agents can be susceptible (S) or infected (I). In this dynamics, at each update, a susceptible individual becomes infected by an infectious neighbor with the probability per unit time $\alpha_0$, and an infected agent recovers with rate $\mu$. In addition, individuals play a public goods game in groups of $G$ individuals. In the public goods game, individuals can be cooperators or defectors. Cooperators pay a cost, $c_G$, to invest in a public good, and defectors pay no cost and do not invest. All the investments in the public good are multiplied by an enhancement factor, $r$, and are divided equally among all the individuals in the group. Thus, each individual $i$ can be in one of four states: $s_i \in \{SC, SD, IC, ID \}$ (Fig. \ref{Fig0}). 
	
	We introduce a coupling between the public goods game and the epidemic dynamics by assuming that cooperation reduces the probability of infection. In order to implement this, we consider both the collective benefit and individual benefit of cooperation in reducing disease spreading by introducing two parameters: altruistic effect, $\alpha_t$, which is defined as the ratio by which the probability that an infected cooperator transmits its disease is reduced, and the self-interested effect, $\alpha_r$, which represents the ratio by which the probability that a susceptible cooperator gets infected is reduced. Thus, the rate that a susceptible cooperator gets infected when in contact with an infected defector is reduced to $\alpha_0\alpha_r$. Similarly, a defector gets infected with rate $\alpha_0\alpha_t$ when in contact with an infected cooperator and with rate $\alpha_0$ when in contact with an infected defector. When both infectious and susceptible agents are cooperators, the disease spreads with rate $\alpha_0\alpha_r\alpha_t$. This procedure is summarized as follows:
	%
	\begin{figure}[t]
		\centering
		\includegraphics[width=.6\linewidth, height=.6\linewidth]{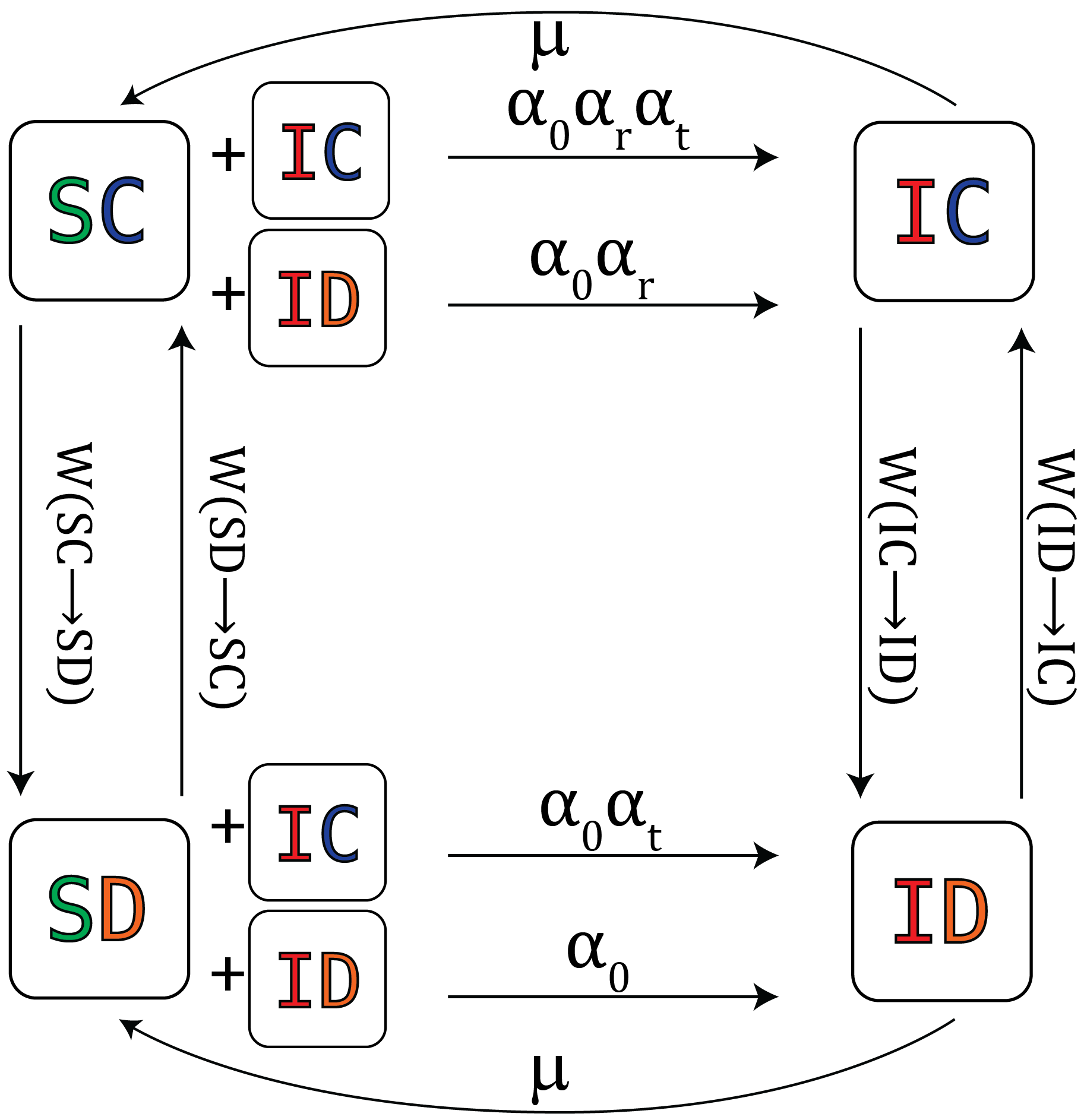}
		\caption{Schematic of model dynamics. The horizontal lines present the infection and recovery processes with given transition rates and the vertical lines show the public goods game.}
		\label{Fig0}
	\end{figure}
	\begin{equation}
		\begin{split}
			SD+ID&\xrightarrow{\hspace{2ex}\alpha_0\hspace{2ex}}ID+ID \\
			SD+IC&\xrightarrow{\hspace{2ex}\alpha_0\alpha_t\hspace{2ex}}ID+IC \\
			SC+IC&\xrightarrow{\hspace{2ex}\alpha_0\alpha_t\alpha_r\hspace{2ex}}IC+IC \\
			SC+ID&\xrightarrow{\hspace{2ex}\alpha_0\alpha_r\hspace{2ex}}IC+ID \\
		\end{split}
		\label{rules}
	\end{equation}
	To improve the realism of our model, we also assume that infected individuals pay an infection cost, $c_I$, which can be related to the treatment or inconveniences caused by getting infected. Thus the net payoffs of an individual in each state $IC$, $ID$, $SC$ and $SD$, and in a group $g$ with $G$ members can be written as follows:
	\begin{equation}
		\begin{split}
			\pi_{IC}^g&=rc_G\frac{N_c+1}{G}-c_G-c_I,\\
			\pi_{ID}^g&=rc_G\frac{N_c}{G}-c_I,\\
			\pi_{SC}^g&=rc_G\frac{N_c+1}{G}-c_G,\\
			\pi_{SD}^g&=rc_G\frac{N_c}{G}.
		\end{split}
		\label{payoff}
	\end{equation}
	Here, $N_c$ is the number of cooperators in the focal individual's group.
	
	Realistically, the spread of the public goods game and the epidemic can occur at different time scales. We denote the characteristic time scale of the game dynamics by $T_{pgg}$, and that of the SIS dynamics by $T_{sis}$. Using these time scales, we can define the probability of playing the game as $\tau=T_{sis}/(T_{sis}+T_{pgg})$. Hence, for $T_{sis} < T_{pgg}$ (small values of $\tau$), the probability of game dynamics is low. For the evolutionary dynamics, firstly, an individual $x$ is randomly selected and plays a public goods game with probability $\tau$ and undergoes disease propagation with probability $1-\tau$. 
	
	For the public goods game dynamics, in a well-mixed population, the focal individual, $x$, plays the PGG in a group of size $G$, with randomly chosen group members. In a structured population, on the other hand, the chosen individuals participate in $n$ public goods game, each centered on one of their neighbors and themselves. The payoff $\Pi_x$ is obtained as the sum of gains acquired from all of the groups in which player $x$ participates during each step. Then, another individual $y$ is randomly selected and its payoff ($\Pi_y$) is similarly calculated. The individual $x$ updates its strategy $s_x$ by adopting the strategy $s_y$ of node $y$, with the following Fermi probability,
	\begin{equation}
		W(s_x\leftarrow s_y)=\frac{1}{1+[\exp(\Pi_x-\Pi_y)/K]},
		\label{fermi}
	\end{equation}
	where, $K$ controls the level of noise. 
	
	If with the probability $1-\tau$, the selected node $x$ is chosen to engage in the epidemic dynamics, it recovers with rate $\mu$ if it is infected and it gets infected by one of infected nodes of the network (in the case of a well-mixed population, $G$ individuals are drawn randomly from the population as neighbors) with a certain rate which depends on the node's states according to the rules of Eq.~\ref{rules}. 
	
	Repeating these elementary steps $N$ times constitutes one full Monte Carlo step (MCS), which ensures that every node has the opportunity to update once on average. The Monte-Carlo steps continue until the dynamics reaches a stationary state (see Table~S1 for a detailed description of simulation).

\section{Methods}
\label{methods}
\subsection{Mean-field description}
		In a well-mixed population, the dynamics of the model can be described using the mean-field method. The mean field equations are obtained by combining the replicator dynamics, commonly used to describe evolutionary dynamics \cite{nowakrepeq}, and a mean-field description of SIS epidemic spreading \cite{pastorsiscomp}. We denote the fraction of individuals in each state $SC$, $SD$, $IC$ and $ID$ by $\rho_{SC}$, $\rho_{SD}$, $\rho_{IC}$ and $\rho_{ID}$, respectively. The mean-field equations are given as follows:
	\begin{widetext}
	\begin{equation}
		\begin{split}
			\frac{d\rho_{SC}}{dt}&=\tau \rho_{SC} (\pi_{SC}-\bar{\pi})
			+(1-\tau)\bigg[ \mu \rho_{IC} - \alpha_0 \alpha_t \alpha_r \rho_{IC} \rho_{SC}
			- \alpha_0 \alpha_r \rho_{ID} \rho_{SC}\bigg],		
			\\
			\frac{d\rho_{SD}}{dt}&=\tau \rho_{SD} (\pi_{SD}-\bar{\pi})
			+ (1-\tau) \bigg[ \mu \rho_{ID} - \alpha_0 \alpha_t \rho_{IC} \rho_{SD}
			- \alpha_0 \rho_{ID} \rho_{SD}\bigg],
			\\
			\frac{d\rho_{IC}}{dt}&=\tau \rho_{IC} (\pi_{IC}-\bar{\pi}) 
			+ (1-\tau) \bigg[ -\mu \rho_{IC} + \alpha_0 \alpha_t \alpha_r \rho_{IC} \rho_{SC} + \alpha_0 \alpha_r \rho_{ID} \rho_{SC}\bigg],
			\\
			\frac{d\rho_{ID}}{dt}&=\tau \rho_{ID} (\pi_{ID}-\bar{\pi}) 
			+(1-\tau) \bigg[ -\mu \rho_{ID} + \alpha_0 \alpha_t \rho_{IC} \rho_{SD}
			+ \alpha_0 \rho_{ID} \rho_{SD}\bigg].
		\end{split}
		\label{mfeq}
	\end{equation}
		\end{widetext}
	In Eq.~\ref{mfeq} with probability $\tau$ the game dynamics occurs and the first terms indicate the replicator dynamics, while the second terms show the dynamics of the epidemic, which occurs with probability  $1-\tau$ in each iteration. Also, $\pi_{SC}$, $\pi_{SD}$, $\pi_{IC}$ and $\pi_{ID}$ are the expected payoffs of individuals of different types and $\bar{\pi}$ is the average payoff of the population, given as follows:
	\begin{equation}
	\begin{split}
		\pi_{SC} &= \sum_{m=0}^{G-1} {G-1\choose m} \rho_C^m\rho_D^{G-1-m}e^{\frac{rc_G\frac{m+1}{G}-c_G}{K}},\\
		\pi_{SD} &= \sum_{m=0}^{G-1} {G-1\choose m} \rho_C^m\rho_D^{G-1-m}e^{\frac{rc_G\frac{m}{G}}{K}},\\
		\pi_{IC} &= \sum_{m=0}^{G-1} {G-1\choose m} \rho_C^m\rho_D^{G-1-m}e^{\frac{rc_G\frac{m+1}{G}-c_G-c_I}{K}},\\
		\pi_{ID} &= \sum_{m=0}^{G-1} {G-1\choose m} \rho_C^m\rho_D^{G-1-m}e^{\frac{rc_G\frac{m}{G}-c_I}{K}},\\
		\bar{\pi} &= \sum_{i}\rho_{i}\pi_{i}, \hspace{10mm} i\in\{IC, ID, SC, SD\}
	\end{split}
	\label{exp-payoff-x}
	\end{equation}
	where, $\rho_{C} \equiv \rho_{SC} + \rho_{IC} $ and $\rho_{D} \equiv \rho_{SD}+ \rho_{ID}$ are the fractions of cooperators and defectors, respectively. 	
	
    To obtain the expected payoffs in Eq.~\ref{exp-payoff-x}, we note that in a group, the focal node plays with $G-1$ neighbors. The probability that there are $m$ cooperators in this group is $\rho_C^m\rho_D^{G-1-m}$, and the binomial coefficient ${G-1\choose m}$, is the number of ways that there are $m$ cooperators among the $G-1$ neighbors of the focal node. We note that as individuals reproduce with a probability proportional to the exponential of their payoff, the expected payoff of a strategy can be defined as the expected value of the exponential of its payoff. Hence for each strategy we multiply by the exponential of the related payoff, and sum over all the possible configurations.

	We solve Eqs.~\ref{mfeq} using the Runge Kutta method and starting from the initial state $\rho_{0SC}=\rho_{0IC}=\rho_{0ID}=\rho_{0SD}=0.25$ until a stationary state is reached. We set $G=5$ and consider the noise to be $K=1.5$ and the recovery rate as $\mu=0.3$. We also set $\alpha_0=1$ in figures where $\alpha_0$ is constant such that the spread of the disease occurs. 
	
\subsection{Simulations}

	We perform numerical simulations for a well-mixed population with size $N = 1000$ and also a square lattice with $L^2=10^4$ nodes and periodic boundary conditions. The initial conditions for the density of cooperator and infected individuals are chosen as $\rho_{0C}=0.5$ and $\rho_{0I}=0.01$. Within the population, cooperators and infected individuals were initially distributed at random. For simplicity, we set the noise in all simulations to be $K=0.5$ and the recovery rate as $\mu=0.3$. 
    
    In the simulation, due to the finite-size effect, $\alpha_0$ depends on $N$. To remove the effect of network size, we rewrite the definition of $\alpha_0$ as the probability per unit time per link. Hence $\alpha_0$ in the simulation is $1/N$ times $\alpha_0$ for the mean-field (well-mixed population). We set $\alpha_0= 0.001$ in numerical simulations for the well-mixed population and $\alpha_0= 0.5$ for the square lattice (in figures where $\alpha_0$ is constant). Furthermore, in all simulations, we assume that the strategy updating is slower than the epidemic spreading, which means $T_{sis} < T_{pgg}$.

    For a well-mixed population, each node plays in a group with $G-1$ randomly chosen individuals ($G=5$). For the disease-spreading process, the focal node can be infected through every infected node in the whole population. However, the square lattice is structured and homogeneous. In this case, in the PGG, the nodes are arranged in overlapping groups with the same size $G=5$. Each player $x$ plays with its $G-1$ partners in the game where it is a focal player and also in the games of his neighbors. The total payoff is then accumulated accordingly. For the disease spreading process on a square lattice, the focal node $x$ can be infected only by the infected nearest neighbors.
    \begin{figure}[t]
    	\centering
    	\includegraphics[width=\linewidth]{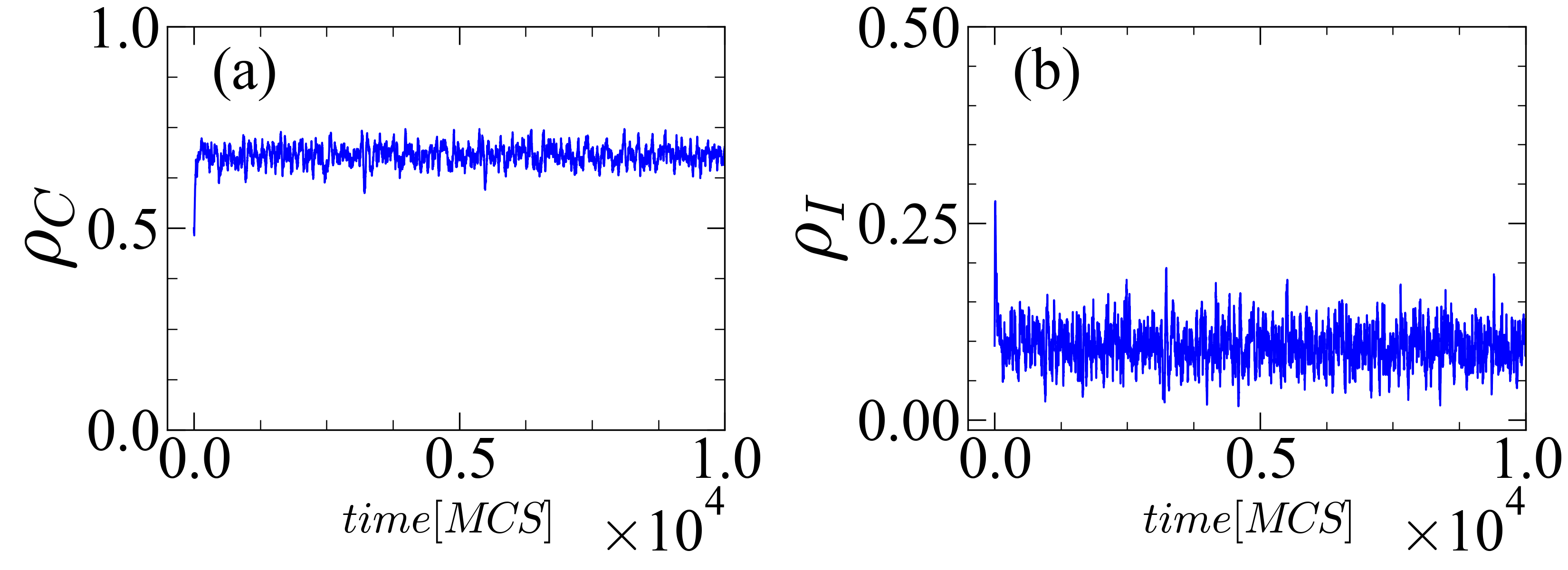}
    	\caption{ $(a)$ Time evolution of the fraction of cooperator and $(b)$ infected individuals in a well-mixed population of size $N=1000$. Parameters values are set as $r=3$, $\alpha_0=0.0016$, $\alpha_r=0.01$, $\alpha_t=0.01$, $c_I=10$, $c_G=1$, and $\tau=0.01$. }
    	\label{Fig1}
    \end{figure}
	\section{Results}
	\label{results}
	 We aim to address how the evolution of cooperation affects the epidemic dynamics, and how the epidemics, in turn, affect the evolutionary dynamics of cooperation. To this end, we focus on the fraction of infected individuals, $\rho_I \equiv \rho_{IC}+ \rho_{ID} $, and the fraction of cooperators, $\rho_C \equiv \rho_{IC}+ \rho_{SC} $. In order to see the effect of disease on cooperation, we set the enhancement factor in a range for which cooperation does not evolve in a simple PGG. Similarly, we set the epidemic dynamics parameters in a range for which epidemics occur in the standard SIS model. Numerical simulations of the model, presented in Fig. \ref{Fig1} show that the presence of infected individuals can drive the evolution of cooperators. Once the cooperators reach a high frequency, the spread of the epidemic is reduced.
	
\begin{figure}[t]
		\centering
		\includegraphics[width=\linewidth]{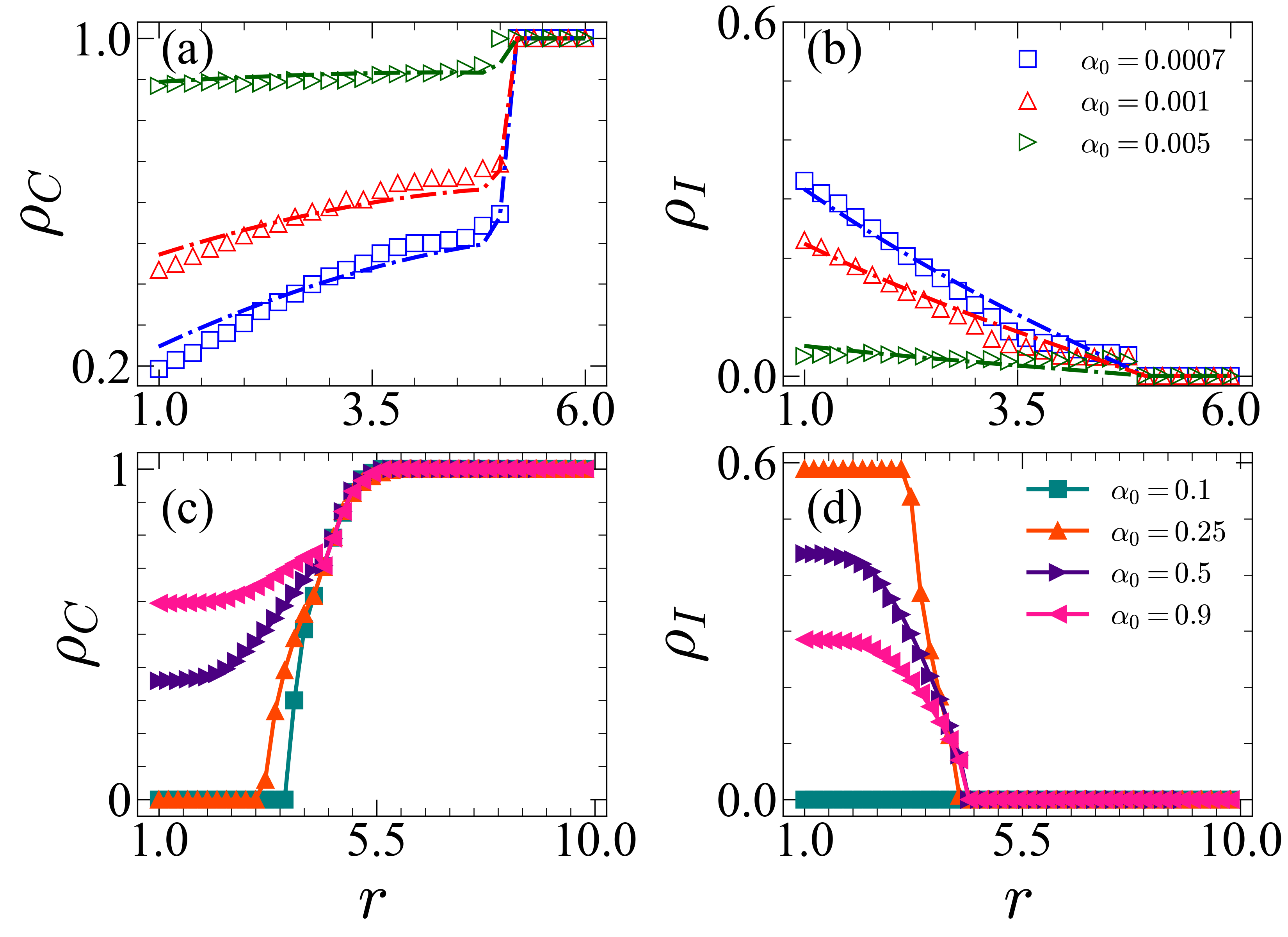}
		\caption{Stationary values for the fraction of cooperators, $\rho_C$, and infected individuals, $\rho_I$, as a function of the enhancement factor, $r$, for different values of infection rate $\alpha_0$ in a well-mixed population of size $N=1000$ ($(a)-(b)$) and a square lattice of size $L^2=10^4$ ($(c)-(d)$). Symbols show numerical simulation results that are in agreement with the mean-field approach (lines) for a well-mixed population. The parameters are set as $\alpha_r=0.01$, $\alpha_t=0.01$, $c_I=10$, $c_G=1$, and $\tau=0.01$. The numerical results are obtained by averaging over 20 realizations.}
		\label{Fig2}
	\end{figure}

	Figure \ref{Fig2}$(a)-(b)$, shows the fraction of cooperator and infected individuals in the stationary state (before the eradication of the epidemics) as a function of the enhancement factor of the PGG, for different values of $\alpha_0$, in a well-mixed population. The results are obtained from both simulations in finite populations, and mean field equations--- in excellent  agreement with simulation results. As we can see the cooperation evolves even for
	$r<G\equiv 5$, where cooperation is a social dilemma and does not evolve in a simple PGG. Surprisingly, higher values of $r$ may only slightly increase cooperation. Rather, they decrease the fraction of the infected individuals. Similarly, in contrast to isolated SIS dynamics, a higher transmission $\alpha_0$ does not lead to a higher fraction of infected individuals. Rather, it leads to higher cooperation, resulting in a more efficient suppression of epidemics. 
	A similar picture is observed in a structured population, presented in Fig. \ref{Fig2}$(c)-(d)$. For too small values of $\alpha_0$, an epidemic does not occur and thus the evolution of cooperation is driven by the evolutionary dynamics of the public goods game. For larger values of $\alpha_0$, the cost of getting infected can drive cooperation even for smaller values of $r$, at which cooperation does not evolve in the absence of an epidemic. For values of $r$ larger than the phase transition point in a standard PGG, the co-evolutionary dynamics is driven by the evolutionary dynamics of PGG. This leads to a reduction of epidemics for high values of $r$. In the square lattice, the results are based on short-range interactions and the correlation between the states of the nodes appears. We provide a description of the spatial correlations in the Supplementary (see Figs. S1, S2 and S3).
	
	\begin{figure}[t]
		\centering
		\includegraphics[width=\linewidth]{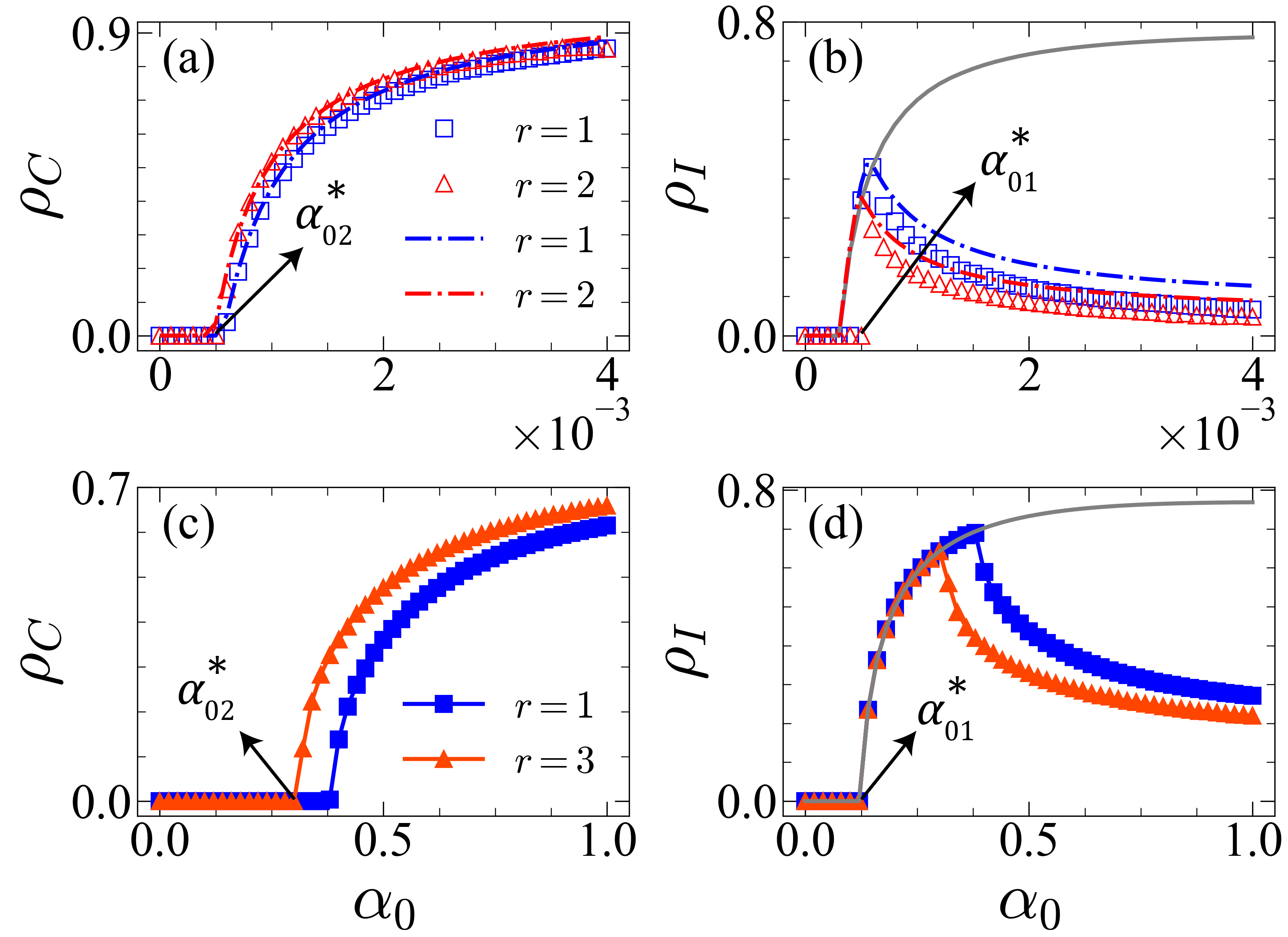}
		\caption{Stationary values for the fraction of cooperators $\rho_C$ and infected nodes $\rho_I$ as a function of infection transmission rate $\alpha_0$ for different values of enhancement factor $r$ in a well-mixed population ($(a)-(b)$) and square lattice ($(c)-(d)$). Transition points to the epidemic state and the cooperative state are presented by $\alpha_{01}^*$ and $\alpha_{02}^*$, respectively. Symbols show numerical simulation results that are in good agreement with the mean-field approach (lines). The solid gray line indicates the fraction of infected individuals in the standard SIS model. Other parameters are set as $\alpha_r=0.01$, $\alpha_t=0.01$, $c_I=10$, $c_G=1$, and $\tau=0.01$. The results are obtained by averaging over 20 realizations.}
		\label{Fig3}
	\end{figure}
	
	To better see how the evolution of cooperation affects the onset of epidemic spreading in the SIS model, we present in Fig. \ref{Fig3} the fraction of cooperators and infected individuals as a function of $\alpha_0$ in a well-mixed and structured population, respectively. In both cases, for too small values of $\alpha_0$, the disease does not spread, and consequently, no cooperation evolves. As $\alpha_0$ increases, we observe a phase transition to an epidemic state at a critical value of $\alpha_{01}^*$. While this transition is sharp in a well-mixed population, it occurs more slowly in a structured population. Once epidemics start to spread, individuals can improve their payoffs by cooperation. Thus, at a second phase transition point, $\alpha_{02}^*$, cooperation evolves. Further increasing $\alpha_0$ beyond $\alpha_{02}^*$, leads to a reduction in the fraction of infected individuals. The extent of disease reduction compared to the density of infected individuals in the standard SIS model is shown in Fig.~\ref{Fig3}$(b)$ and $(d)$ (represented by the solid gray line).
	
\begin{figure}[t]
			\centering
			\includegraphics[width=\linewidth]{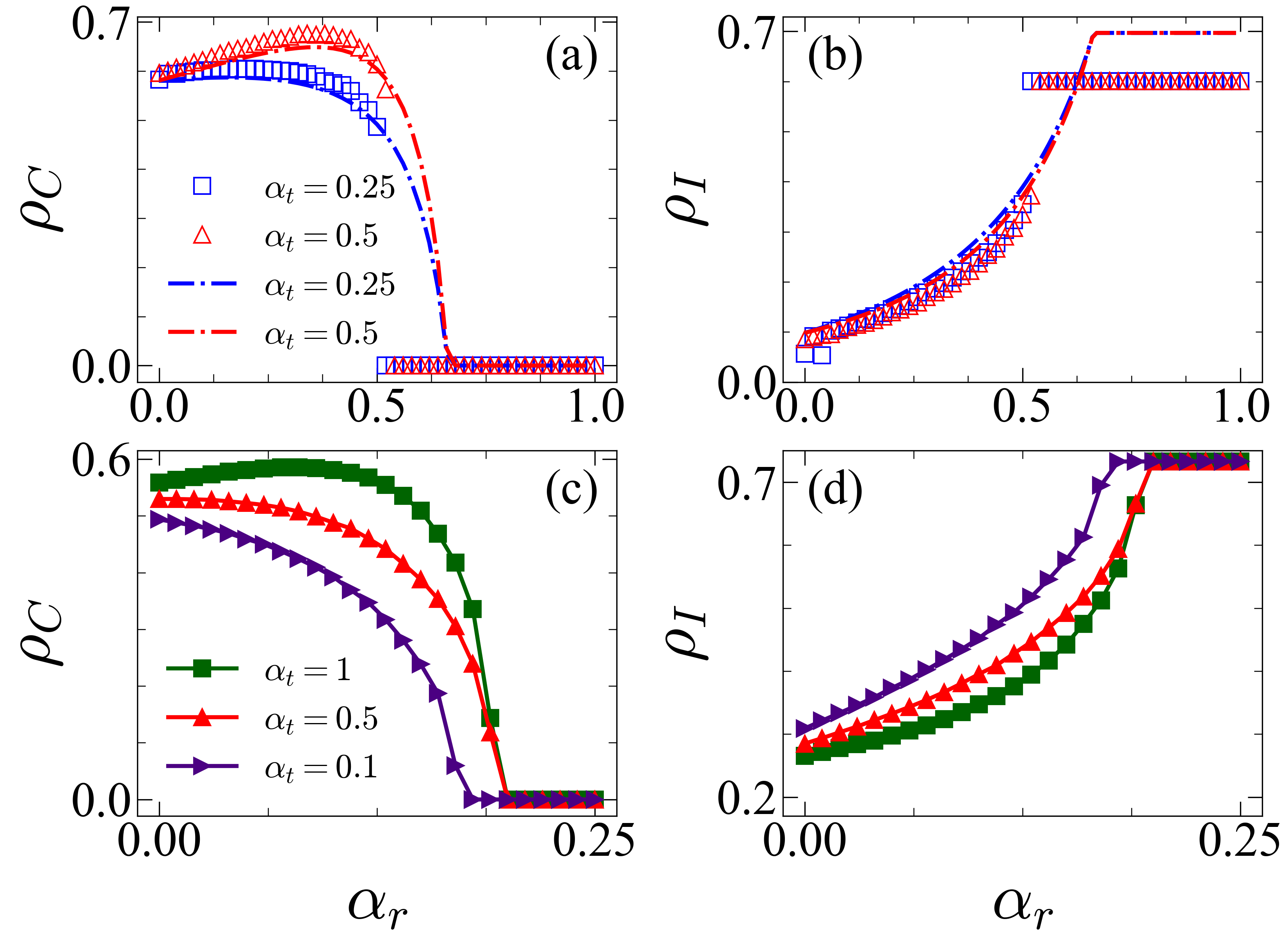}	
			\caption{The density of cooperators $\rho_C$ and infected individuals $\rho_I$ as a function of self-interested protection effect $\alpha_r$ in a well-mixed population ($(a)-(b)$) and square lattice ($(c)-(d)$). Different curves are related to different values of altruistic protection effect $\alpha_t$. Symbols show numerical simulation results which are in rather good agreement with the mean-field approach (lines) for a well-mixed population. Other parameters are set as $r=3$, $c_I=10$, $c_G=1$, and $\tau=0.01$. The results are obtained by averaging over 20 realizations.}
			\label{Fig4}
		\end{figure}

	An important question is what factors underlie the evolution of cooperation in the face of an epidemic. In our model, cooperation has two effects on the epidemic dynamics. Firstly, through self-interested protective behavior, it reduces the risk of getting sick. Secondly, it also has an altruistic effect by reducing the probability of transmission of the disease to others. To investigate which factor underlies cooperation, in Fig.~\ref{Fig4}, we present $\rho_C$ and $\rho_I$ as a function of the self-interested protection effect $\alpha_r$, for different values of altruistic protection effect, $\alpha_t$. For high values of $\alpha_r$, cooperation is not effective in reducing the probability of getting infected. Therefore, cooperation does not evolve. As $\alpha_r$ decreases, cooperation becomes more efficient and evolves below a critical value of $\alpha_r$. 
		
		In a well-mixed population, the altruistic protection parameter has a small, yet negative effect on cooperation. That is, for smaller $\alpha_t$ (higher altruistic protection efficiency), cooperation decreases more. This is due to the fact that higher altruistic protection efficiency more strongly controls the epidemics, and thus, the cost of cooperation can outweigh the benefit of cooperation in avoiding infection.
		
		A similar, yet stronger trend is observed in a structured population; higher self-interested protection efficiency (smaller $\alpha_r$) promotes cooperation and curbs disease transmission. In contrast, surprisingly, higher altruistic protection efficiency (smaller $\alpha_t$) decreases cooperation and increases infection. This is due to the fact that in a structured population, higher altruistic benefit of cooperation leads to a healthier local environment, and thus, a reduced net benefit of cooperation for the individual. These results show that, counter-intuitively, the collective benefit of cooperation in controlling the epidemics can only adversely affect disease control. Furthermore, in contrast to a simple PGG where population structure is beneficial for the evolution of cooperation, population structure can be detrimental to altruistic behavior in controlling an epidemic.
		\begin{figure}[t]
			\centering
			\includegraphics[width=\linewidth]{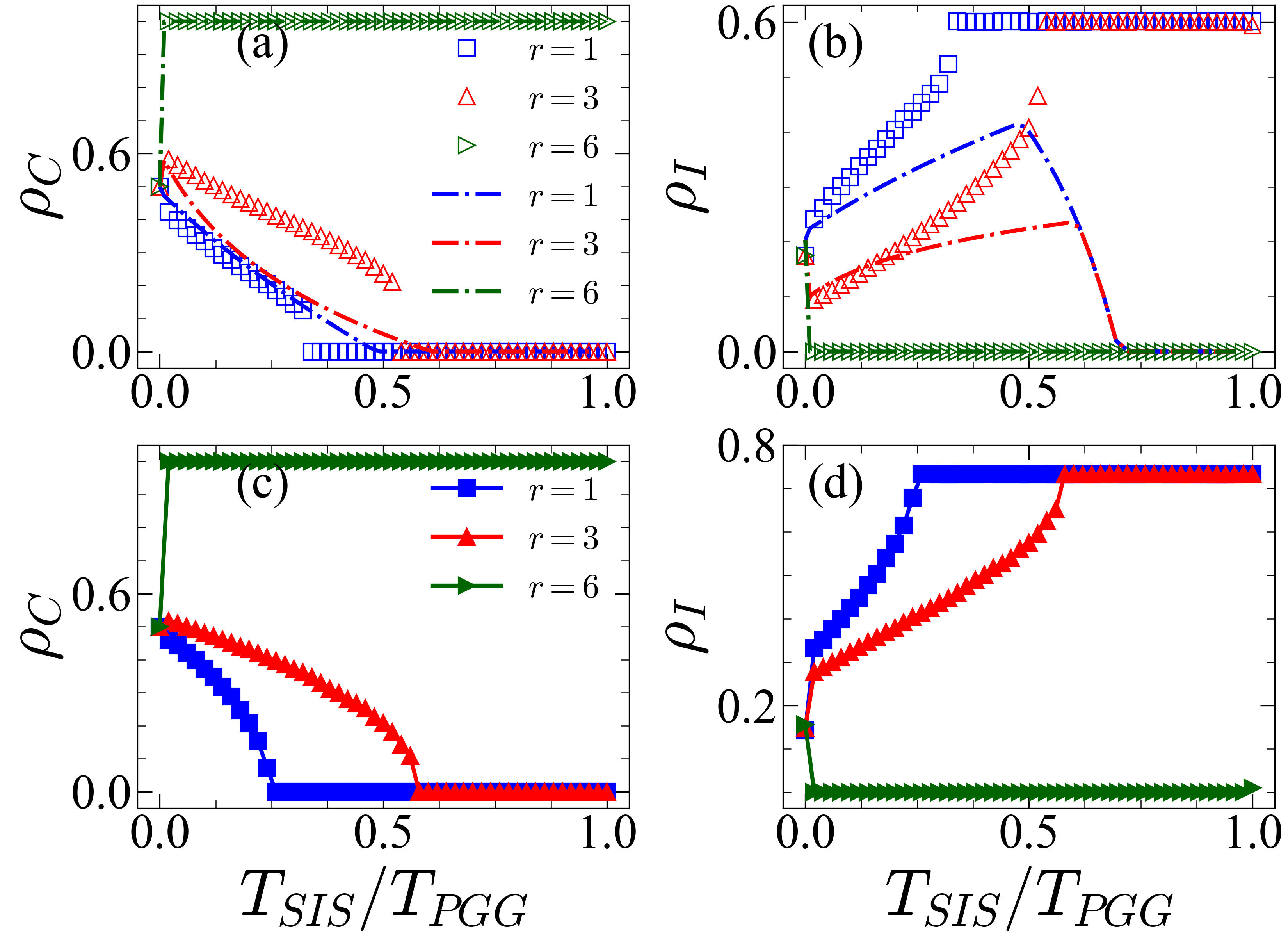}
			\caption{The density of cooperators $\rho_C$ and infected individuals $\rho_I$ as a function of time-scale ratio $T_{sis}/T_{pgg}$ for different values of enhancement factor $r$ in a well-mixed population ($(a)-(b)$) and a square lattice ($(c)-(d)$). Symbols show numerical simulation results and for well-mixed populations, the lines indicate the mean-field solutions. Other parameters are set as $\alpha_r=0.01$, $\alpha_t=0.01$, $c_I=10$, and  $c_G=1 $. The results are obtained by averaging over 20 realizations.}
			\label{Fig5}
		\end{figure}
		
		So far we have considered the coevolution of two dynamics of the SIS epidemic spreading and the PGG for small values of $\tau$. In other words, we assumed that the disease spreading is much faster than game evolution, which seems to be a realistic assumption in many contexts, such as when the cooperative dynamics is a genetic trait, or individuals only slowly change their behavior by learning. In Fig. \ref{Fig5}, we investigate the effect of the relative speed of the two dynamics on the emergence of cooperation and the spread of disease, by plotting the fraction of cooperator and infected individuals as a function of the ratio of the time scales of the two dynamics. First, we consider $r < r^*$, both for well-mixed population and the square lattice, such that cooperation does not exist in the standard PGG. The simulation results show that for large values of $T_{sis}/T_{pgg}$, since the disease spreads at a low speed, there is little benefit of cooperation in controlling the epidemics and the dynamics is predominantly governed by the evolutionary dynamics of PGG. As a result, eventually, the disease covers a large percentage of the population. However, by increasing the speed of disease (low $\tau$), cooperation emerges and leads to the reduction of the epidemic. As we can see, there is a discrepancy between the simulation results and the mean field solutions, especially at high values of $T_{sis}/T_{pgg}$ in the mixed populations. We can explain this discrepancy as follows: for large values of $T_{sis}/T_{pgg}$, the dynamics of the game is faster, leading to a rapid decline in the number of cooperators. Consequently, the influence of the first and third equations from the set of equations in Eq.~\ref{mfeq} diminishes, leaving only the second and fourth equations, which describe the evolution of $\rho_{SD}$ and $\rho_{ID}$, to dictate the dynamics. In these remaining equations, the first term related to replicator dynamics exerts a more substantial effect. Given that infected defectors have a lower payoff ($\pi_{ID} <\pi_{SD} $), the population of infected individuals gradually declines. As illustrated in Fig.~\ref{Fig5}, at high values of $T_{sis}/T_{pgg}$, the infected population approaches zero, which contradicts the simulation outcomes. Therefore, it seems that the mean field equations are in good agreement with the simulation results only for low values $T_{sis}/T_{pgg}$ (low $\tau$). Moreover, the general reason for the discrepancy can be related to the fact that in the Monte Carlo simulations, time is discretized into time steps of length $\Delta t$ (typically $\Delta t = 1$), and events occur with certain probabilities rather than transition rates between the states in the continuous-time mean field equations. The probabilities are simply the product of the corresponding rates and the time step $\Delta t$. The agreement between the discrete-time Monte Carlo simulations and the continuous-time mean-field method increases at $\Delta t \to 0$, meaning that the transition probabilities must be chosen very small \cite{fennellLimitations}. 	
		
		While we have focused on the regime where PGG is a social dilemma ($r<5$), as can be seen in Fig. \ref{Fig5}, for $r=6$, not surprisingly, cooperation can easily evolve and eradicate the epidemics, provided the basis controlling effect of cooperation is strong enough (small enough $\alpha_r$ and $\alpha_t$).
		\begin{figure}[t]
			\centering
			\includegraphics[width=\linewidth]{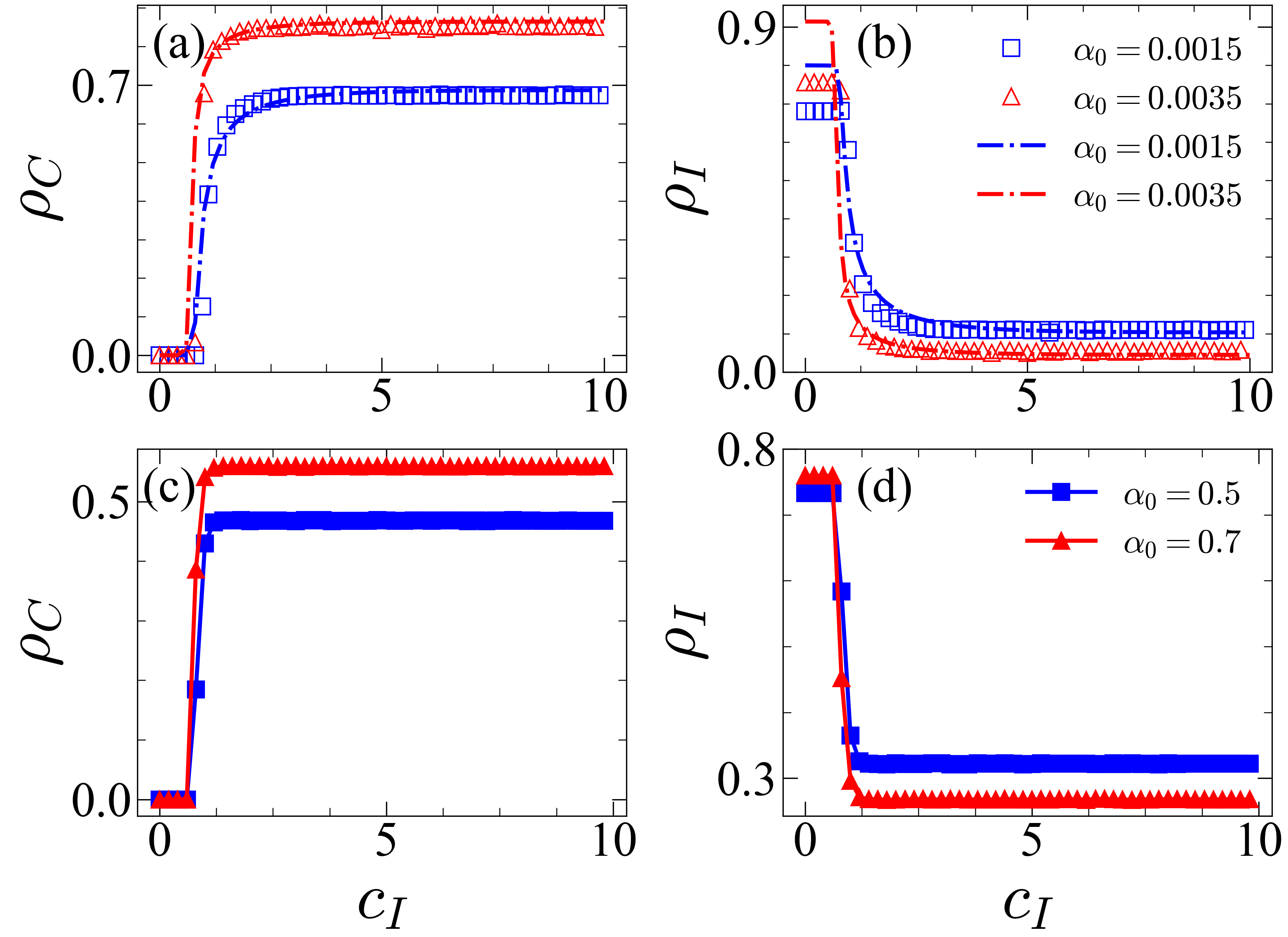}	
			\caption{ The density of cooperators $\rho_C$ and infected individuals $\rho_I$ as a function of infection cost $c_I$ for two values of infection transmission rate $\alpha_0$ in a $(a)-(b)$ well-mixed population and $(c)-(d)$ square lattice. Symbols show numerical simulation results and for well-mixed population the lines indicate the mean-field solutions. Other parameters are set as $\alpha_r=0.01, \alpha_t=0.01, r=2, \tau=0.01$. The results are obtained by averaging over 10 realizations.}
			\label{Fig6}
		\end{figure}
			
		Another important parameter of the model is the cost of infection, $c_I$, which affects the coupling of the epidemics and evolutionary dynamics by changing the payoffs of individuals depending on their infected status. $\rho_C$ and $\rho_I$ as a function of the cost of infection in a well-mixed and structured population, and for two different values of $\alpha_0$ are presented in Fig.~\ref{Fig6}. Cooperation can reduce the probability of infection for individuals. Thus, the expected payoff of cooperators increases by increasing the cost of infection, and cooperation evolves only for $c_I$ above a phase transition threshold, where infection is costly enough for the individuals to motivate cooperation. Consequently, high levels of cooperation and a reduction of infection is observed only for high values of $c_I$.	
		
\begin{figure}[t]
		\centering
		\includegraphics[width=\linewidth]{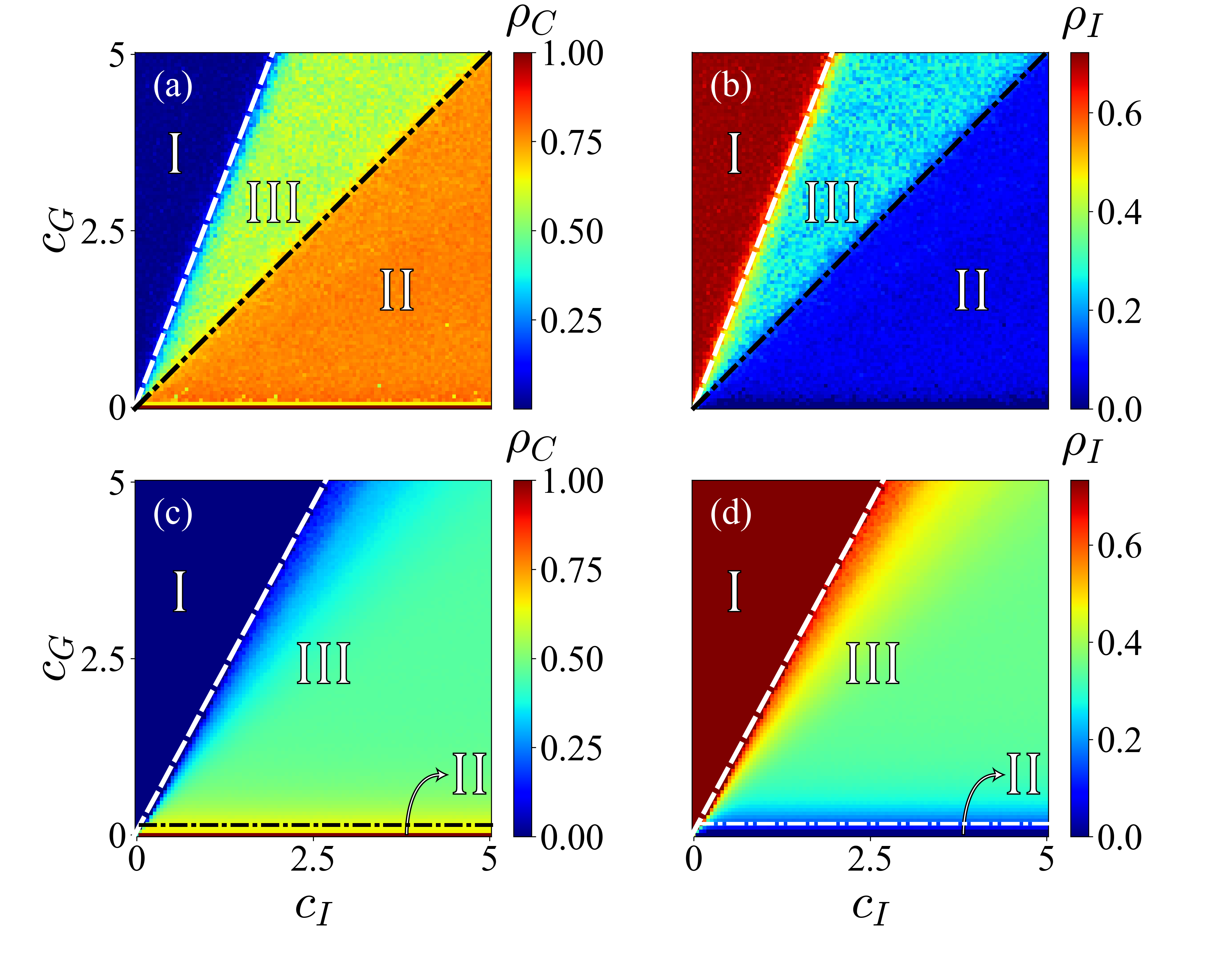}
		\caption{Phase diagram of the model for values of $\rho_C$ and $\rho_I$ as functions of infection cost $c_I$ and cooperation cost $c_G$ in a  well-mixed population $((a)-(b))$ and square lattice $((c)-(d))$. The other parameters are set as $r=3, \alpha_r=0.01, \alpha_t=0.01$, and $\tau=0.01$.}
		\label{Fig7}
	\end{figure}

		The effect of costs on the cooperation level and epidemics can be explored further. We show the phase diagram for the model in the space $c_G -c_I$ in Fig.~\ref{Fig7}. We can see different regimes, namely,
		
		\begin{itemize}
			\item non-cooperative state (I): wherein a large number of nodes is infected while all the nodes are defector ($\rho_{C}=0$). This happens above the white dashed line.
			\item disease-free state (II): all nodes are susceptible and also some nodes are cooperator ($\rho_{I}=0$). This happens below the dashed-dotted line.
			\item epidemic and cooperative state (III): a fraction of the nodes is infected and also some nodes are cooperator.
		\end{itemize}
	\begin{figure}[t]
		\centering
		\includegraphics[width=1.1\linewidth, height=0.8\linewidth]{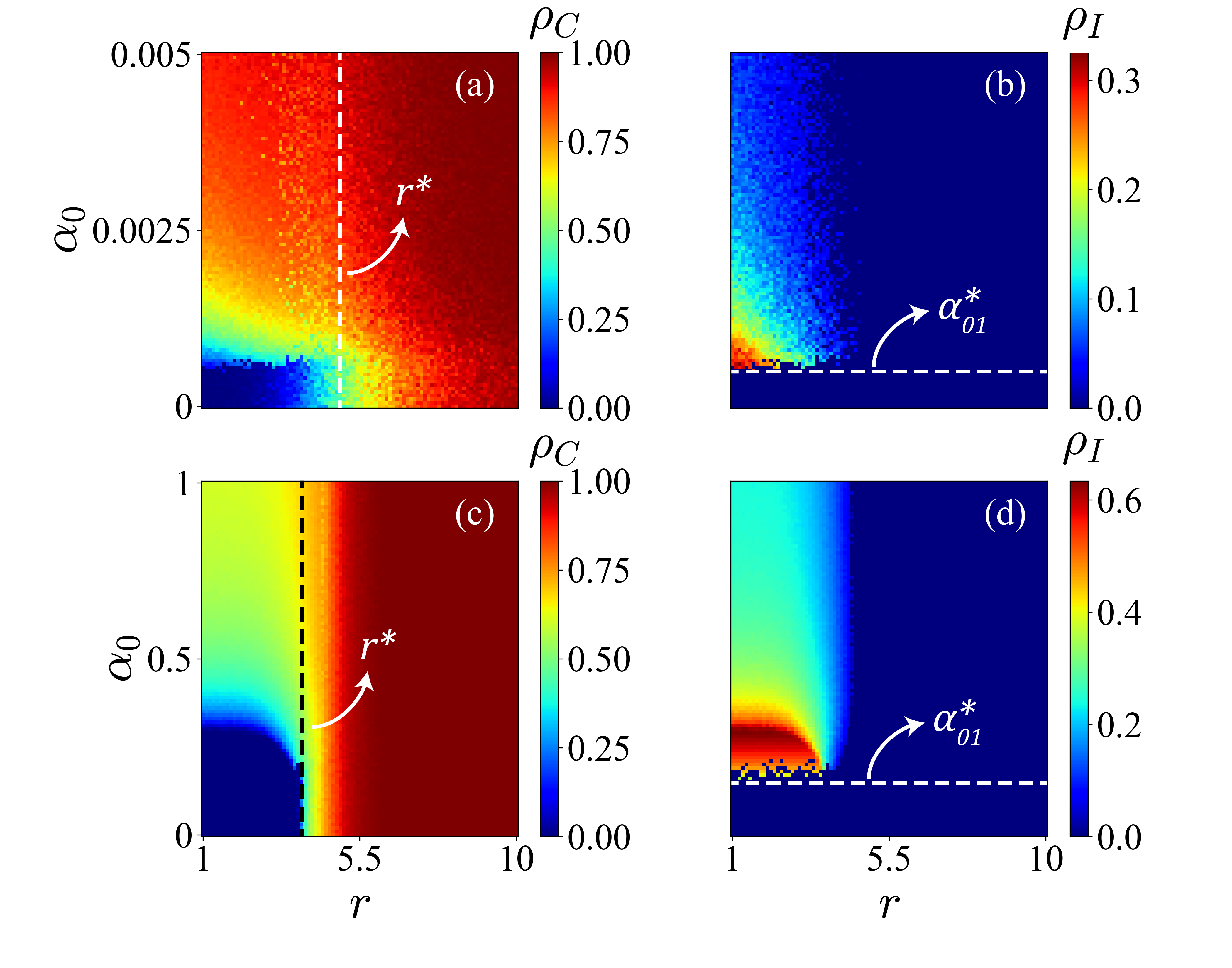}
		\caption{Phase diagram of the model for values of $\rho_C$ and $\rho_I$ as functions of enhancement factor $r$ and infection transmission rate $\alpha_0$ in a well-mixed population $((a)-(b))$ and square lattice $((c)-(d))$. The other parameters are set as $\alpha_r=0.1, \alpha_t=0.5, c_I=10, c_G=1$, and $\tau=0.01$.}
		\label{Fig8}
	\end{figure}
		
		According to Fig.~\ref{Fig7}, for cooperation to exist in the system, the cost of infection $c_I$ must be greater than the cost of cooperation $c_G$. Also, it is observed that with the increase in the cost of infection, the spread of disease decreases. In the case $c_G \approx 0$, cooperation is at its maximum and the disease is completely eradicated.
		
		In Fig.~\ref{Fig8}, we can see the level of cooperation and extent of the epidemics for different enhancement factors $r$ and disease transmission probability $\alpha_0$. For the values of $\alpha_0 < \alpha_{01} ^*$  there is no disease in the system and cooperation occurs above the threshold point $r^*$, as it is for the standard PGG. With increasing $\alpha_{0}$, which signifies a higher prevalence of the disease in the system, the enhancement threshold for the onset of cooperation gradually decreases. This illustrates how the existence of the disease affects the enhancement of cooperation. With increasing $r$ and the emergence of cooperation, the epidemic decreases. For large values of $r$, where the cooperation has encompassed the entire system, the disease completely disappears.
		
		\section{Conclusion}
		\label{conclusion}
		We have studied the interplay between the SIS epidemic model and the public goods game, by considering a context where cooperation can reduce the probability of infection through both self-interested and altruistic protective measures and in turn, infection can affect evolutionary dynamics by imposing a cost on infected individuals. Our findings reveal that the presence of an epidemic can promote cooperation even in regimes where cooperation does not evolve in the standard PGG. We found that the coupling of the two dynamics can modify the phenomenology and phase transitions in both simple PGG and simple SIS models. Higher epidemic transmissions can promote the evolution of cooperation in PGG, leading to reduced prevalence of infection in the population. Similarly, a higher benefit of cooperation in the public goods game can positively impact the control of the epidemic dynamics by facilitating cooperation.
		
		Moreover, we observed that the self-interest parameter $\alpha_r$ and the altruism parameter $\alpha_t$ play crucial roles in shaping the dynamics of cooperation and infection. While higher self-interested protection efficiency (lower values of $\alpha_r$) promotes cooperation due to its positive impact on the individuals' health, higher altruistic protection efficiency (lower values of $\alpha_t$), surprisingly, can undermine cooperation by controlling the epidemics, and thus reducing the benefit of cooperation in keeping the individuals in a healthy state. Furthermore, our results indicate that the homogeneous nature of square lattices facilitates the clustering of cooperators, which enhances local herd immunity and suppresses the epidemic more effectively than in well-mixed populations. However, this effect can hinder the evolution of altruistic behavior in controlling a disease in a structured population.

		Our study highlights the importance of considering both epidemiological and behavioral factors in understanding and controlling infectious diseases. Future research could extend this framework to more heterogeneous contexts, such as complex networks, varying recovery rates, or adaptive network typologies to better capture the complex dynamics resulting from the coupling of evolutionary dynamics and disease spreading.
		
\section{Acknowledgement}
M.S. Acknowledges funding from Deutsche Forschungsgemeinschaft (German Research Foundation) under Germany's Excellence Strategy-EXC 2117-422037984.
		

		\bibliographystyle{ieeetr}
		\bibliography{refs}

\begin{thebibliography}{10}

\bibitem{bogua2003}
M.~Bogu{\'a}, R.~Pastor-Satorras, and A.~Vespignani, {\em Statistical Mechanics
  of Complex Networks}, ch.~Epidemic Spreading in Complex Networks with Degree
  Correlations.
\newblock Springer Berlin Heidelberg, 2003.

\bibitem{wang2024}
W.~Wang, Y.~Nie, W.~Li, T.~Lin, M.-S. Shang, S.~Su, Y.~Tang, Y.-C. Zhang, and
  G.-Q. Sun, ``Epidemic spreading on higher-order networks,'' {\em Physics
  Reports}, vol.~1056, pp.~1--70, 2024.

\bibitem{koher2019}
A.~Koher, H.~H.~K. Lentz, J.~P. Gleeson, and P.~H\"ovel, ``Contact-based model
  for epidemic spreading on temporal networks,'' {\em Phys. Rev. X}, vol.~9,
  p.~031017, 2019.

\bibitem{zhou2019}
Y.~Zhou, J.~Zhou, G.~Chen, and H.~E. Stanley, ``Effective degree theory for
  awareness and epidemic spreading on multiplex networks,'' {\em New Journal of
  Physics}, vol.~21, no.~3, p.~035002, 2019.

\bibitem{moez2006}
M.~Draief, ``Epidemic processes on complex networks the effect of topology on
  the spread of epidemics,'' {\em Physica A: Statistical Mechanics and its
  Applications}, vol.~363, no.~1, pp.~120--131, 2006.

\bibitem{newman2002}
M.~E.~J. Newman, ``Spread of epidemic disease on networks,'' {\em Phys. Rev.
  E}, vol.~66, no.~11, p.~016128, 2002.

\bibitem{pastor2015}
R.~Pastor-Satorras, C.~Castellano, P.~Van~Mieghem, and A.~Vespignani,
  ``Epidemic processes in complex networks,'' {\em Rev. Mod. Phys.}, vol.~87,
  no.~55, pp.~925--979, 2015.

\bibitem{wang2017}
W.~Wang, M.~Tang, H.~E. Stanley, and L.~A. Braunstein, ``Unification of
  theoretical approaches for epidemic spreading on complex networks,'' {\em
  Reports on Progress in Physics}, vol.~80, no.~3, p.~036603, 2017.

\bibitem{wang2016}
Z.~Wang, C.~T. Bauch, S.~Bhattacharyya, A.~d'Onofrio, P.~Manfredi, M.~Perc,
  N.~Perra, M.~Salathé, and D.~Zhao, ``Statistical physics of vaccination,''
  {\em Physics Reports}, vol.~664, pp.~1--113, 2016.

\bibitem{Han}
H.~Chuanliang, L.~Meijia, H.~Naem, {\em et~al.}, ``Mechanisms of recurrent
  outbreak of covid-19: a model-based study,'' {\em Nonlinear Dyn}, vol.~106,
  p.~1169–1185, 2021.

\bibitem{Claudio}
R.~P.-S. C.~Castellano, ``Competing activation mechanisms in epidemics on
  networks,'' {\em Sci Rep}, vol.~2, p.~371, 2012.

\bibitem{Bavel}
V.~Bavel, J.~J, Baicker, Katherine, Boggio, P.~S, {\em et~al.}, ``Using social
  and behavioural science to support covid-19 pandemic response.,'' {\em Nat
  Hum Behav}, vol.~4, pp.~460--471, 2020.

\bibitem{pastor2002}
R.~Pastor-Satorras and A.~Vespignani, ``Immunization of complex networks,''
  {\em Phys. Rev. E}, vol.~65, p.~036104, 2002.

\bibitem{schneider2011}
C.~M. Schneider, T.~Mihaljev, S.~Havlin, and H.~J. Herrmann, ``Suppressing
  epidemics with a limited amount of immunization units,'' {\em Phys. Rev. E},
  vol.~84, p.~061911, 2011.

\bibitem{khanjani2020}
M.~Khanjanianpak, N.~Azimi-Tafreshi, and C.~Castellano, ``Competition between
  vaccination and disease spreading,'' {\em Phys. Rev. E}, vol.~101, p.~062306,
  2020.

\bibitem{gosak2021community}
M.~Gosak, M.~Duh, R.~Markovič, and M.~Perc, ``Community lockdowns in social
  networks hardly mitigate epidemic spreading,'' {\em New Journal of Physics},
  vol.~23, no.~4, p.~043039, 2021.

\bibitem{gosak2021endogenous}
M.~Gosak, M.~U.~G. Kraemer, H.~H. Nax, M.~Perc, and B.~S.~R. Pradelski,
  ``Endogenous social distancing and its underappreciated impact on the
  epidemic curve,'' {\em Sci. Rep.}, vol.~11, no.~1, p.~3093, 2021.

\bibitem{SMH1}
S.~L. Loo, M.~Chinazzi, A.~Srivastava, S.~Venkatramanan, S.~Truelove, and
  C.~Viboud, ``Preface: Covid-19 scenario modeling hubs,'' {\em Epidemics},
  vol.~48, p.~100788, 2024.

\bibitem{SMH2}
R.~K. Borchering, J.~M. Healy, B.~L. Cadwell, M.~A. Johansson, R.~B. Slayton,
  M.~Wallace, and M.~Biggerstaff, ``Public health impact of the u.s. scenario
  modeling hub,'' {\em Epidemics}, vol.~44, p.~100705, 2023.

\bibitem{morison2024}
C.~Morison, M.~Fic, T.~Marcou, J.~Mohamadichamgavi, J.~Redondo~Antón,
  G.~Sayyar, A.~Stein, F.~Bastian, H.~Krakovská, N.~Krishnan, D.~Pires, M.~R.
  Satouri, F.~J. Thomsen, K.~Tjikundi, and W.~Ali, ``Public goods games in
  disease evolution and spread,'' 2024.

\bibitem{dees2018}
R.~H. Dees, ``Public health and normative public goods,'' {\em Public Health
  Ethics}, vol.~11, no.~1, pp.~20--26, 2018.

\bibitem{yong2021}
J.~C. Yong and B.~K.~C. Choy, ``Noncompliance with safety guidelines as a
  free-riding strategy: An evolutionary game-theoretic approach to cooperation
  during the covid-19 pandemic,'' {\em Frontiers in Psychology}, vol.~12, 2021.

\bibitem{soltan2020}
M.~Soltanolkottabi, D.~Ben-Arieh, and C.-H. Wu, ``Game theoretic modeling of
  infectious disease transmission with delayed emergence of symptoms,'' {\em
  Games}, vol.~11, no.~2, 2020.

\bibitem{tanimoto2015wp}
J.~Tanimoto, {\em Fundamentals of Evolutionary Game Theory and its
  Applications}.
\newblock Springer, 2015.

\bibitem{chang2020ug}
S.~L. Chang, M.~Piraveenan, P.~Pattison, and M.~Prokopenko, ``Game theoretic
  modelling of infectious disease dynamics and intervention methods: a
  review,'' {\em J. Biol. Dyn.}, vol.~14, no.~1, pp.~57--89, 2020.

\bibitem{huang2022qh}
Y.~Huang and Q.~Zhu, ``Game-theoretic frameworks for epidemic spreading and
  human decision-making: A review,'' {\em Dyn. Games Appl.}, vol.~12, no.~1,
  pp.~7--48, 2022.

\bibitem{amaral2021}
M.~A. Amaral, M.~M. de~Oliveira, and M.~A. Javarone, ``An epidemiological model
  with voluntary quarantine strategies governed by evolutionary game
  dynamics,'' {\em Chaos, Solitons and Fractals}, vol.~143, p.~110616, 2021.

\bibitem{hamilton}
A.~Hamilton, F.~Haghpanah, A.~Tulchinsky, N.~Kipshidze, S.~Poleon, G.~Lin,
  H.~Du, L.~Gardner, and E.~Klein, ``Incorporating endogenous human behavior in
  models of covid-19 transmission: A systematic scoping review,'' {\em
  Dialogues Health.}, vol.~4, p.~100179, 2024.

\bibitem{wang2020uc}
Z.~Wang and C.~Xia, ``Co-evolution spreading of multiple information and
  epidemics on two-layered networks under the influence of mass media,'' {\em
  Nonlinear Dyn.}, vol.~102, no.~4, pp.~3039--3052, 2020.

\bibitem{wang2015coupled}
Z.~Wang, M.~A. Andrews, Z.-X. Wu, L.~Wang, and C.~T. Bauch, ``Coupled
  disease--behavior dynamics on complex networks: A review,'' {\em Physics of
  Life Reviews}, vol.~15, pp.~1--29, 2015.

\bibitem{wang2020vaccination}
X.~Wang, D.~Jia, S.~Gao, C.~Xia, X.~Li, and Z.~Wang, ``Vaccination behavior by
  coupling the epidemic spreading with the human decision under the game
  theory,'' {\em Applied Mathematics and Computation}, vol.~380, p.~125232,
  2020.

\bibitem{khanjanianpak2022ap}
M.~Khanjanianpak, N.~Azimi-Tafreshi, A.~Arenas, and J.~G{\'o}mez-Garde{\~n}es,
  ``Emergence of protective behaviour under different risk perceptions to
  disease spreading,'' {\em Phil. Trans. R. Soc. A.}, vol.~380, no.~2227,
  p.~20200412, 2022.

\bibitem{haurent2002}
C.~Hauert, S.~{De Monte}, J.~Hofbauer, and K.~Sigmund, ``Volunteering as red
  queen mechanism for cooperation in public goods games,'' {\em Science},
  vol.~296, no.~5570, pp.~1129--1132, 2002.

\bibitem{hauert2003wc}
C.~Hauert and G.~Szab{\'o}, ``Prisoner's dilemma and public goods games in
  different geometries: Compulsory versus voluntary interactions,'' {\em
  Complexity}, vol.~8, no.~4, pp.~31--38, 2003.

\bibitem{hauert2008ecological}
C.~Hauert, J.~Y. Wakano, and M.~Doebeli, ``Ecological public goods games:
  Cooperation and bifurcation,'' {\em Theoretical Population Biology}, vol.~73,
  no.~2, pp.~257--263, 2008.

\bibitem{santos2008social}
F.~C. Santos, M.~D. Santos, and J.~M. Pacheco, ``Social diversity promotes the
  emergence of cooperation in public goods games,'' {\em Nature}, vol.~454,
  no.~7201, pp.~213--216, 2008.

\bibitem{xianbin2010}
X.-B. Cao, W.-B. Du, and Z.-H. Rong, ``The evolutionary public goods game on
  scale-free networks with heterogeneous investment,'' {\em Physica A:
  Statistical Mechanics and its Applications}, vol.~389, no.~6, pp.~1273--1280,
  2010.

\bibitem{perc2017}
M.~Perc, J.~J. Jordan, D.~G. Rand, Z.~Wang, S.~Boccaletti, and A.~Szolnoki,
  ``Statistical physics of human cooperation,'' {\em Physics Reports},
  vol.~687, pp.~1--51, 2017.

\bibitem{Zhu2024}
Z.~Wenqiang, Wang, Xin, Wang, Chaoqian, {\em et~al.}, ``Evolutionary dynamics
  in stochastic nonlinear public goods games,'' {\em Commun Phys}, vol.~7,
  p.~377, 2024.

\bibitem{Tim2020}
T.~Johnson, C.~T. Dawes, J.~H. Fowler, and O.~Smirnov, ``Slowing covid-19
  transmission as a social dilemma: Lessons for government officials
  frominterdisciplinary research on cooperation,'' {\em Journal of Behavioral
  Public Administration}, vol.~3, 2020.

\bibitem{nowakrepeq}
M.~A. Nowak, {\em Evolutionary Dynamics: Exploring the Equations of Life}.
\newblock Harvard University Press, 2006.

\bibitem{pastorsiscomp}
R.~Pastor-Satorras and A.~Vespignani, ``Epidemic dynamics and endemic states in
  complex networks,'' {\em Phys. Rev. E}, vol.~63, p.~066117, 2001.

\bibitem{fennellLimitations}
P.~G. Fennell, S.~Melnik, and J.~P. Gleeson, ``Limitations of discrete-time
  approaches to continuous-time contagion dynamics,'' {\em Phys. Rev. E},
  vol.~94, p.~052125, 2016.

\end{thebibliography}

	\end{document}